\documentclass{iopart}
\usepackage{graphicx}
\usepackage{amssymb}
\usepackage{html}
\usepackage{url}
\usepackage{harvard}

\newcommand{\qudit}[1]{\left\vert #1 \right\rangle}

\newcommand{\Z}{\mathbb{Z}}

\newcommand{\C}{\mathbb{C}}

\newcommand{\II}{\mathbb{I}}
\newtheorem{thm}{Theorem}[section]

\newtheorem{lemma}[thm]{Lemma}

\newtheorem{Problem}[thm]{Problem}
\begin{document}
\bibliographystyle{jphysicsB}
\title{On swapping the states of two qudits}

\author{Colin M Wilmott}
\address{Institut f\"ur Theortische Physik III,  Heinrich-Heine-Universit\"at, D\"usseldorf, Germany}
\ead{wilmott@thphy.uni-duesseldorf.de}
\date{Received: date / Revised version: date}

\begin{abstract}
The {{SWAP}} gate has become an integral feature of quantum
circuit architectures and is designed to permute the states of two
qubits through the use of the well-known controlled-NOT gate. We
consider the question of whether a two-qudit quantum circuit
composed entirely from instances of the generalised
controlled-{{NOT}} gate can be constructed to permute the states
of two qudits. Arguing via the signature of a permutation, we
demonstrate the impossibility of such circuits for dimensions $d
\equiv 3$ (mod $4$).
\end{abstract}

\section{Introduction}

A prerequisite for quantum computation is the successful
implementation of multiple-qubit quantum gates. The most
elementary of all multiple-qubit quantum gates is premised by
two-qubit controlled unitary operators, and a classic example is
the controlled-{\small{NOT}} ({\small{CNOT}}) gate. The
{\small{CNOT}} gate has been shown to provide a basis for a
measurement set that permits the construction of  syndrome tables
used in error correction \cite{PhysRevA.55.114}. Additionally,
the {\small{CNOT}} gate has assumed a central  role in the theory
of quantum computation. It is the quantum mechanical analogue of
the classical connective {\small{XOR}} gate and is a principle
component in universal computation. Furthermore,
\citeasnoun{BBCDM95} have shown that any multiple-qubit quantum
operation may be restricted to compositions of single-qubit gates
and instances of the {\small{CNOT}} gate. It is for this reason
that we say the quantum gate library consisting of single-qubit
gates and the {\small{CNOT}} gate is universal. As a consequence,
the {\small{CNOT}} gate has acquired the special status as the
hallmark of multi-qubit control \cite{Vidaldawson04}.

In recent years, researchers in universal quantum computation have
done considerable work optimizing quantum circuit constructions.
 \citeasnoun{VW04} constructed a quantum circuit for general
two-qubit operations which requires fifteen single-qubit gates and
at most three {\small{CNOT}} gates. A crucial aspect of this
result is the demand that the {\small{SWAP}} gate requires at
least three {\small{CNOT}} gates. The {\small{SWAP}} gate
describes the cyclical permutation of the states of two qubits and
has become an integral feature of the circuitry design for many
quantum operators. It is a fundamental element in the circuit
implementation of Shor's algorithm \cite{FDH04}, and
\citeasnoun{LL05} maintain that experimentally realising the
 {\small{SWAP}} gate is a necessary condition for the
networkability of quantum computation.

In this paper, we examine the possibility of constructing a
two-qudit {\small SWAP} gate using only instances of the
generalised {\small CNOT} gate to permute the states of two
qudits. Section \ref{pre} introduces preliminary material from the
theory of permutations which will serve as a basis for our study.
Section \ref{d=3} considers the particular problem of whether a
two-qutrit quantum circuit composed entirely from instances of the
two-qutrit {\small CNOT} gate can be constructed to permute the
states of two qutrits. Finally, section \ref{d=d} generalises the
results of section \ref{d=3} to two-qudit quantum systems before
demonstrating the impossibility of two-qudit {\small SWAP} gates
using only instances of the generalised {\small CNOT} gate for
dimensions $d \equiv 3$ (mod $4$).

\section{Preliminaries}\label{pre}

\begin{figure}
\setlength{\unitlength}{0.08cm} \hspace*{95mm} \hskip-4.5em
\begin{picture}(40,40)(68.4,0)
\put(-25,30){(a)} \put(0,20){\line(1,0){20}}
\put(0,30){\line(1,0){20}} \put(10,18){\line(0,1){11}}
\put(10,20){\circle{4}} \put(10,30){\circle*{2}}
\put(-12,29){${\tiny{\qudit{\psi}_{{{{}}}}}}$}
\put(-12,19){${\tiny{\qudit{\phi}_{{{{}}}}}}$}
\put(24,29){${\tiny{\qudit{\psi}_{{{{}}}}}}$}
\put(24,19){${\tiny{\qudit{\phi\oplus \psi}}}$}
\end{picture}
\begin{picture}(40,25)(-55,0)
\put(-25,30){(b)} \put(0,20){\line(1,0){20}}
\put(0,30){\line(1,0){20}} \put(10,20){\line(0,1){12}}
\put(10,20){\circle*{2}} \put(10,30){\circle{4}}
\put(-12,29){${\tiny{\qudit{\psi{{\cal{{}}}}}}}$}
\put(-12,19){${\qudit{\phi}}$} \put(24,29){${\tiny{\qudit{\psi
\oplus \phi}_{{\cal{{}}}}}}$}
\put(24,19){${\tiny{\qudit{\phi}_{{\cal{{}}}}}}$}
\end{picture}
\vskip-3em\caption{Circuit descriptions for the {{CNOT}} gate
types. (a) The  {{CNOT1}} gate; the state of control system
$\qudit{\psi} \in  {\cal H}_{\cal A}$ remains unchanged after
application  whereas the state of the target system $\qudit{\phi}
\in {\cal H}_{\cal B}$ is transformed under modular arithmetic to
the state $\qudit{\phi\oplus \psi}$. (b) The {{CNOT2}} gate in
which the roles of systems ${\cal H}_{\cal A}$ and ${\cal H}_{\cal
B}$ are reversed. }\label{quditCNOT}
\end{figure}
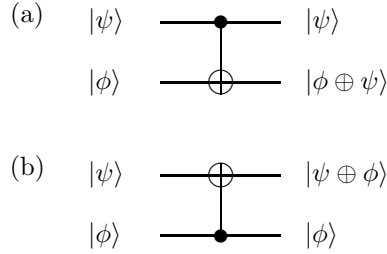

\begin{figure}
\setlength{\unitlength}{0.08cm} \hspace*{95mm} \hskip-4.5em
\begin{picture}(40,40)(63,0)
\put(0,20){\line(1,0){40}} \put(0,30){\line(1,0){40}}
\put(10,18){\line(0,1){11}} \put(10,20){\circle{4}}
\put(10,30){\circle*{2}} \put(20,20){\line(0,1){12}}
\put(20,20){\circle*{2}} \put(20,30){\circle{4}}
\put(30,18){\line(0,1){11}} \put(30,20){\circle{4}}
\put(30,30){\circle*{2}}
\put(-7,29){$\qudit{\psi}$} \put(-7,19){$\qudit{\phi}$}
\put(44,29){$\qudit{\phi}$} \put(44,19){$\qudit{\psi}$}
\end{picture}
\vskip-3em\caption{The  {\small{SWAP}} gate illustrating the
cyclical permutation of two qubits. System ${\mathcal{A}}$ begins
in the state $\qudit{\psi}$ and ends in the state $\qudit{\phi}$
while system ${\mathcal{B}}$ begins in the state $\qudit{\phi}$
and ends in the state $\qudit{\psi}$.}\label{swap}
\end{figure}
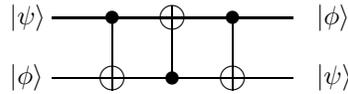

\subsection{Elementary quantum gates}
Let $\cal H$ denote the $d$-dimensional complex Hilbert space
$\C^d$. Fix each orthonormal basis state of the $d$-dimensional
Hilbert  space to correspond to an element of $\Z_d$; as such the
basis $\{\qudit{0},\qudit{1},\dots,\qudit{d-1}\} \subset \C^d$
whose elements correspond to the column vectors of the  identity
matrix ${\II}_d$ is called the computational basis. A \emph{qudit}
is a $d$-dimensional quantum state $\qudit{\psi} \in {\cal H}$
written as  $\qudit{\psi} = \sum^{d-1}_{i=0}{\alpha_i\qudit{i}}$
where $\alpha_i \in \C$ and
$\sum^{d-1}_{i=0}{\vert\alpha_i\vert^2} = 1$. Given
$d$-dimensional Hilbert spaces ${\cal H}_{\cal A}$ and ${\cal
H}_{\cal B}$, consider the set of $d^2 \times d^2$ unitary
transformations $U \in {U(d}^2)$ that act on the two-qudit quantum
system ${\cal H}_{\cal A} \otimes {\cal H}_{\cal B}$. Let
$U_{\textrm{\tiny CNOT1}} \in U(d^2)$ represent the generalised
{\small{CNOT}} gate that has control qudit  $\qudit{\psi} \in
{\cal H}_{\cal A}$ and  target  qudit $\qudit{\phi} \in {\cal
H}_{\cal B}$. The action of $U_{\textrm{\tiny CNOT1}}$ on the set
of basis states $\qudit{m}\otimes\qudit{n}$ of ${\cal H}_{\cal A}
\otimes {\cal H}_{\cal B}$ is  given by
\begin{eqnarray}
U_{\textrm {\tiny
 CNOT1}}\qudit{m}_{}\otimes\qudit{n}_{} =
\qudit{m}_{}\otimes\qudit{n \oplus m}_{}, \qquad m,n\in \Z_d,
\end{eqnarray}
with $\oplus$ denoting addition  modulo $d$. Similarly, let
$U_{\textrm {\tiny CNOT2}} \in U(d^2)$   denote the generalised
{\small{CNOT}} gate having  control qudit $\qudit{\phi}_{} \in
{\cal H}_{\cal B}$ and target qudit $\qudit{\psi}_{} \in {\cal
H}_{\cal A}$. The action of $U_{\textrm{\tiny CNOT2}}$ on the set
of basis states $\qudit{m}_{}\otimes\qudit{n}_{}$ of  ${\cal
H}_{\cal A} \otimes {\cal H}_{\cal B}$ is written
\begin{eqnarray}
U_{\textrm{\tiny{CNOT2}}}\qudit{m}\otimes\qudit{n}_{{}} =
\qudit{m\oplus n}\otimes\qudit{n}_{{}}, \qquad m,n\in \Z_d.
\end{eqnarray}
Figure~\ref{quditCNOT} provides the quantum gate circuitry
representation  for the respective {\small{CNOT}} types while
figure~\ref{swap} illustrates the well-known  {\small SWAP} gate
for permuting the states of two qubits.

\subsection{Permutation groups}
Consider the set $\textrm{N} = \{1,2,\dots,n\}$ and let
$\sigma: \textrm{N} \mapsto \textrm{N}$ be a bijection. Let
$\sigma =
\left[%
\begin{array}{cccc}
 1&2&\dots&n \\
  i_1&i_2&\dots&i_n \\
\end{array}%
\right]$ be a \emph{permutation} of the set $\textrm{N}$ with $i_k
\in \textrm{N}$ denoting the image of $k \in \textrm{N}$ under
$\sigma$. Let $\sigma$ and $\tau$ be two permutations of
$\textrm{N}$. Define the product $\sigma\cdot\tau$ by
$(\sigma\cdot\tau)(i) = \sigma(\tau(i))$,  $i \in \textrm{N}$, to
be the composition of the mapping $\tau$ followed by $\sigma$.
These permutations taken with $(\cdot)$ form the group $S_n$
called  the \emph{symmetric group} of degree $n$.

Given the permutation $\sigma \in S_n$ and for each $i \in
\textrm{N}$, let us consider the sequence $i, \sigma(i),
\sigma^2(i), \dots.$ Since $\sigma$ is a bijection and
$\textrm{N}$ is finite there exist a smallest positive integer
$\ell = \ell(i)$ depending on $i$  such that $\sigma^\ell(i) = i.$
The \emph{orbit} of $i$ under $\sigma$ then consists of  the
elements $i, \sigma(i),\dots,\sigma^{\ell-1}(i)$. By a
\emph{cycle} of $\sigma$, we mean the ordered set $(i,
\sigma(i),\dots,\sigma^{\ell-1}(i))$ which sends $i$ into
$\sigma(i)$, $\sigma(i)$ into $\sigma^2(i)$,$ \dots,
\sigma^{\ell-2}(i)$ into $\sigma^{\ell-1}(i)$, and
$\sigma^{\ell-1}(i)$ into  $i$ and leaves all other elements of
\textrm{N} fixed. Such a cycle is called an $\ell$-cycle. We refer
to $2$-cycles as \emph{transpositions} and note that any
permutation can be written as a product of transpositions. A pair
of elements $\{\sigma(i), \sigma(j)\}$ is  an  \emph{inversion} in
a permutation $\sigma$ if $i < j$ and $\sigma(i)>\sigma(j)$. The
number of transpositions in any such product is even if and only
if the number of inversions is even. Consequently, we say such a
permutation is even. A similar case holds for  odd permutations.

\begin{lemma}\label{decomposition}
Every permutation can be uniquely expressed as a product of
disjoint cycles.
\end{lemma}
\noindent{\bf{Proof.}} Let $\sigma$ be a permutation. Then the
cycles of the permutation are of the form $i,
\sigma(i),\dots,\sigma^{\ell-1}(i)$. Since the cycles are disjoint
and by the multiplication of cycles, we have it that the image of
$i \in N$ under $\sigma$ is the same as the image under the
product, $\varsigma$, of all the disjoint cycles of $\sigma$. Then
$\sigma$ and $\varsigma$ have the same effect on every element in
$N$, hence, $\sigma = \varsigma$. \hfill $\Box$

Every permutation  $\sigma \in S_n$ has  a \emph{cycle
decomposition} that is unique up to the ordering of the cycles and
up to a cyclic permutation of the elements within each cycle.
Further, if $\sigma \in S_n$ and $\sigma$ is written as the
product of disjoint cycles of length $n_1,\dots,n_k$, with
$n_i\leq n_{i+1}$, we say $(n_1,\dots,n_k)$ is the \emph{cycle
type} of $\sigma$. As a result of Lemma \ref{decomposition}, every
permutation can be written as a product of transpositions. Since
the number of transpositions needed to represent a given
permutation is either even or odd, we  define the \emph{signature}
of a permutation as
\begin{eqnarray}
\textrm{sgn}(\sigma) = \left\{%
\begin{array}{cc}
  +1 & \textrm{if $\sigma$ is even} \\
  -1  & \textrm{if $\sigma$ is odd.} \\
\end{array}%
\right.
\end{eqnarray}
To each permutation $\sigma \in S_n$,  let us consider the
corresponding
permutation matrix $A_{\sigma}$ whereby \begin{eqnarray} A_{\sigma}{(j,i)} = \left\{%
\begin{array}{cc}
  1 & \textrm{if $\sigma(i)$ = $j$} \\
  0 & \textrm{otherwise.} \\
\end{array}
\right.
\end{eqnarray}
The mapping $f: S_n \mapsto \textrm{det}(A_\sigma)$  where
\begin{eqnarray}
\textrm{det}(A_\sigma) = \sum_{\sigma\in
S_n}{\textrm{sgn}(\sigma)\prod_{i=1}^{n}{A_{\sigma(i),i}}}
\end{eqnarray}
is a group homomorphism.  The kernel of this homomorphism,
$\textrm{ker}f$, is  the set of even permutations. Consequently,
we have it that $\sigma \in S_n$ is even if and only if
$\textrm{det}(A_\sigma)$ equals $+1$.

\section{On swapping the states of two qutrits}\label{d=3}


Let $d=3$ and  consider the following problem. Given a pair of
qutrit quantum systems, system ${\cal{A}}$ in the state
$\qudit{\psi}$ and system ${\cal{B}}$ in the state $\qudit{\phi}$,
and using only instances of the two-qutrit {\small CNOT} gate,
determine if it is possible permute the states of the
corresponding systems so that system ${\cal{A}}$ ends in the state
$\qudit{\phi}$ while system ${\cal{B}}$ ends in the state
$\qudit{\psi}$.

\begin{Problem}\label{Problem_Qutrit}
Given qutrits $\qudit{\psi} \in {\cal H}_{\cal A}$ and
$\qudit{\phi} \in {\cal H}_{\cal B}$ and using only instances of
the two-qutrit {\small CNOT} gate, determine if it is possible to
construct a two-qutrit quantum circuit that permutes the states of
the quantum systems ${\cal H}_{\cal A}$ and ${\cal H}_{\cal B}$
such that $\qudit{\psi}\otimes\qudit{\phi} \in {\cal H}_{\cal
A}\otimes{\cal H}_{\cal B}$ is mapped to
$\qudit{\phi}\otimes\qudit{\psi} \in {\cal H}_{\cal A}\otimes{\cal
H}_{\cal B}$.

\end{Problem}
\begin{figure}
\begin{picture}(85,10)(-90,70)
\put(-21,43){(a)}


\put(0,10){$\left(%
\begin{array}{ccccccccc}
  1 & 0 & 0 & 0 & 0 & 0 & 0 & 0 & 0 \\
  0 & 1 & 0 & 0 & 0 & 0 & 0 & 0 & 0 \\
  0 & 0 & 1 & 0 & 0 & 0 & 0 & 0 & 0 \\
  0 & 0 & 0 & 0 & 0 & 1 & 0 & 0 & 0 \\
  0 & 0 & 0 & 1 & 0 & 0 & 0 & 0 & 0 \\
  0 & 0 & 0 & 0 & 1 & 0 & 0 & 0 & 0 \\
  0 & 0 & 0 & 0 & 0 & 0 & 0 & 1 & 0 \\
  0 & 0 & 0 & 0 & 0 & 0 & 0 & 0 & 1 \\
  0 & 0 & 0 & 0 & 0 & 0 & 1 & 0 & 0 \\
\end{array}\right)$}%
\end{picture}

\begin{picture}(85,250)(-90,-50)
\put(-21,43){(b)}
\put(0,10){$\left(%
\begin{array}{ccccccccc}
  1 & 0 & 0 & 0 & 0 & 0 & 0 & 0 & 0 \\
  0 & 0 & 0 & 0 & 0 & 0 & 0 & 1 & 0 \\
  0 & 0 & 0 & 0 & 0 & 1 & 0 & 0 & 0 \\
  0 & 0 & 0 & 1 & 0 & 0 & 0 & 0 & 0 \\
  0 & 1 & 0 & 0 & 0 & 0 & 0 & 0 & 0 \\
  0 & 0 & 0 & 0 & 0 & 0 & 0 & 0 & 1 \\
  0 & 0 & 0 & 0 & 0 & 0 & 1 & 0 & 0 \\
  0 & 0 & 0 & 0 & 1 & 0 & 0 & 0 & 0 \\
  0 & 0 & 1 & 0 & 0 & 0 & 0 & 0 & 0 \\
\end{array}%
\right)$}
\end{picture}
\caption{Matrix representations of two-qutrit {{CNOT}} types. (a)
The matrix representation for the two-qutrit {{CNOT1}} gate. (b)
The matrix description for the two-qutrit {{CNOT2}} gate.}
\label{3CNOT}
\end{figure}

We now show that for a  pair of qutrits, it is not possible to
permute the states  $\qudit{\psi}\otimes\qudit{\phi} \in {\cal
H}_{\cal A}\otimes{\cal H}_{\cal B}$ using only instances of the
two-qutrit {\small CNOT} gate.

Any two-qutrit quantum circuit  composed entirely from instances
of the two-qutrit {\small CNOT} gate can be written in terms of
the two-qutrit {\small{CNOT1}} and {\small{CNOT2}} gates. The
action of the two-qutrit {\small{CNOT1}} gate on the basis states
$\qudit{m}_{}\otimes\qudit{n}_{} \in {\cal H}_{\cal A} \otimes
{\cal H}_{\cal B}, m,n \in \Z_3,$ is described by the unitary
transformation $U_{\textrm {\tiny
 CNOT1}} \in U(9)$ given by
\begin{eqnarray}
U_{\textrm {\tiny CNOT1}}\qudit{m}_{}\otimes\qudit{n}_{} =
\qudit{m}_{}\otimes\qudit{n \oplus m}_{}, \qquad m, n \in \Z_3,
\end{eqnarray}
where $\oplus$ denotes addition  modulo $3$. Figure \ref{3CNOT}
(a) provides the matrix  description for the two-qutrit {\small
CNOT1} gate. A similar description  for  the two-qutrit
{\small{CNOT2}} gate holds, and figure \ref{3CNOT} (b) provides
the corresponding matrix description. Note also that the
two-qutrit {\small{CNOT1}} and {\small{CNOT2}} gates can be
described in the following way. The permutation matrix
corresponding to the two-qutrit {\small{CNOT1}} gate takes the
value 1 in row $3m+n$ and column $3m + (n\ominus m),$   $m,n \in
\Z_3.$ Similarly, the matrix corresponding to the two-qutrit
{\small{CNOT2}} gate takes the value 1 in row $3m+n$ column
$3(m\ominus n) + n$, $m,n \in \Z_3$. Importantly, both of these
matrix descriptions
 have determinant +1 as  the permutations corresponding to their
the respective matrices are even.

\begin{figure}
\begin{picture}(95,120)(-30,0)
\put(60,50){$\left(%
\begin{array}{ccccccccc}
  1 & 0 & 0 & 0 & 0 & 0 & 0 & 0 & 0 \\
  0 & 0 & 0 & 1 & 0 & 0 & 0 & 0 & 0 \\
  0 & 0 & 0 & 0 & 0 & 0 & 1 & 0 & 0 \\
  0 & 1 & 0 & 0 & 0 & 0 & 0 & 0 & 0 \\
  0 & 0 & 0 & 0 & 1 & 0 & 0 & 0 & 0 \\
  0 & 0 & 0 & 0 & 0 & 0 & 0 & 1 & 0 \\
  0 & 0 & 1 & 0 & 0 & 0 & 0 & 0 & 0 \\
  0 & 0 & 0 & 0 & 0 & 1 & 0 & 0 & 0 \\
  0 & 0 & 0 & 0 & 0 & 0 & 0 & 0 & 1 \\
\end{array}%
\right)$}\end{picture} \caption{The matrix $U^* \in U(9)$ that
permutes the states of two qutrits. }\label{3interchange}
\end{figure}

Let us now assume there exists a two-qutrit quantum circuit
composed entirely from instances of the two-qutrit {\small{CNOT}}
gate types which permutes the states of two qutrits. By
assumption, such a circuit will then be a composition of
two-qutrit {\small{CNOT1}} and {\small{CNOT2}} gates. It then
follows that any composition of two-qutrit {\small{CNOT1}} and
{\small{CNOT2}} gates will be equivalent to some product of their
respective unitary matrix descriptions. Such a  matrix product
will necessarily have determinant +1 as both constituent elements
 have determinant +1. However, figure \ref{3interchange}
represents the unitary transformation $U^* \in U(9)$ required to
permute  the states of two qutrits. Such a swap matrix takes the
value 1 in row $3m+n$ column $3n + m$, and has determinant $-1$.
Therefore, no composition of two-qutrit {\small CNOT} gate types
can yield the required matrix, and the result follows. \hfill
$\Box$

\section{On swapping the states of two qudits}\label{d=d}
 We  generalize   problem \ref{Problem_Qutrit} to
higher dimensional quantum systems and ask if it is possible to
construct a two-qudit quantum circuit  composed entirely from
instances of the generalised {\small CNOT} gate to  permute the
states of two qudit quantum systems.

\begin{Problem}\label{Problem_Qudit}
Given  a pair of qudits $\qudit{\psi} \in {\cal H}_{\cal A}$ and
$\qudit{\phi} \in {\cal H}_{\cal B}$ and using only instances of
the generalised {\small CNOT} gate, determine if it is possible to
construct a two-qudit quantum circuit to permute the state
$\qudit{\psi}\otimes\qudit{\phi} \in {\cal H}_{\cal A}\otimes{\cal
H}_{\cal B}$.
\end{Problem}

We have shown in section \ref{d=3} that  the unitary matrices
corresponding to the two-qutrit {\small{{CNOT1}}} and
{\small{{CNOT2}}} gate types both have determinant +1, and this
contrasted significantly with the determinant of the unitary
matrix required to permute the states of a pair of qutrits.
Consequently, no composition of the former could yield the latter
and the result followed. There is, however, another way to look
the problem of permuting the states of two quantum systems using
only instances of the generalised {\small CNOT} gate, and it is
the following. Firstly, in examining the qutrit case, we note that
the permutations corresponding to the two-qutrit {CNOT1} matrix
and the swap matrix $U^* \in U(9)$ are
\begin{eqnarray}\sigma_{\textrm{\tiny{CNOT1}}} &=&
\left[%
\begin{array}{ccccccccc}
  0 & 1 & 2 & 3 & 4 & 5 & 6 & 7 & 8 \\
  0 & 1 & 2 & 5 & 3 & 4 & 7 & 8 & 6 \\
\end{array}%
\right]\\
\sigma_{{{\tiny{U}}}^*} &=& \left[%
\begin{array}{ccccccccc}
  0 & 1 & 2 & 3 & 4 & 5 & 6 & 7 & 8 \\
  0 & 5 & 6 & 7 & 4 & 1 & 2 & 3 & 8 \\
\end{array}%
\right]\end{eqnarray}  respectively. In particular, these
permutations describe, respectively, the action of both the
two-qutrit {CNOT1} gate and the swap matrix $U^*$ on the set of
basis states $\qudit{m}\otimes\qudit{n}\in
{\cal{H}}_{\cal{{A}}}\otimes{\cal{H}}_{\cal{{B}}},\  m,n \in\Z_3$.
The cycle type for two-qutrit {\small{CNOT}} gate is $(1,1,1,3,3)$
while the cycle type for the swap matrix $U^*$ is $(1,1,1,2,2,2)$.
Hence, the two-qutrit {\small{CNOT}} gate fixes three basis states
and permutes the remaining states in two cycles of length 3. Each
such cycle may be written as a product of two transpositions.
Whence, the signature of  the two-qutrit {\small{CNOT}}
permutation is +1. On the other hand, the permutation describing
swap of a pair of qutrit states contains three fixed elements and
a set of three transpositions and therefore the signature of this
permutation is $-1$. Consequently, it follows that within a
two-qutrit quantum circuit architecture, no composition of the
two-qutrit {\small{CNOT}} gate types alone can permute the states
of two qutrits.

More generally, the two-qudit {\small{CNOT}} gate that acts on a
pair of $d$-dimensional quantum systems corresponds to a
permutation of the $d^2$ basis states.  For prime dimensions
$d=p$, the permutation associated with the generalised
{\small{CNOT1}} gate fixes $d$ basis states and induces $(d-1)$
cycles of length $d$, each of which may be written as a product of
$(d-1)$ transpositions. The generalised {\small{CNOT1}} gate
permutation is then a composition of $(d-1)^2$ basis state
transpositions. A similar case holds for the generalised
{\small{CNOT2}} gate in that corresponding mapping fixes $d$ basis
elements induces $(d-1)$ cycles where each is a product of $(d-1)$
transpositions. Therefore, the  signature of the generalised
{\small{CNOT}} permutation is $-1$ for dimension $d=2$ and $+1$
for odd prime dimensions.

Now, let us consider the unitary matrix $U^* \in U(d^2)$ that
swaps the states of two qudits. This matrix permutes the basis
states of a pair of qudits thereby mapping the two-qudit state
$\qudit{\psi}\otimes \qudit{\phi}\in {\cal{H}}_{\cal{A}} \otimes
{\cal{H}}_{\cal{B}}$ to the state $\qudit{\phi}\otimes
\qudit{\psi} \in {\cal{H}}_{\cal{A}} \otimes {\cal{H}}_{\cal{B}}$.
Such a transformation corresponds to a permutation of the $d^2$
basis states $\qudit{m}\otimes \qudit{n} \in {\cal{H}}_{\cal{A}}
\otimes {\cal{H}}_{\cal{B}}$, $m,n \in \Z_d$.  Under this mapping,
there are $d$ fixed basis elements and $d(d-1)/2$ transpositions
on all remaining basis states. Consequently, the signature of the
permutation describing the two-qudit {\small SWAP} gate is $-1$
for dimensions $d \equiv 2$ or $3$ (mod $4$) and $+1$ for
dimensions $d \equiv 0$ or $1$ (mod $4$). It then follows that
two-qudit quantum circuits composed entirely from instances of the
generalised {\small{CNOT}} gate can not permute the states  of two
 qudits when $d \equiv 3$ (mod $4$). \hfill $\Box$

\section{Conclusion}\label{conclusion}

We considered the problem of constructing a two-qudit {\small
SWAP} gate  using only instances of the generalized {\small CNOT}
gate. We discussed the idea of signature for a permutation and
identified, via this signature, when it is possible to permute the
states of two qudits using only instances of the generalised
{\small CNOT} gate. Based on this argument, we demonstrated the
impossibility of constructing a two-qudit {\small SWAP} gate using
only instances of the generalised {\small CNOT} gate for
dimensions $d \equiv 3$ (mod $4$). This may be of interest the for
more general task of constructing a $k$-qudit {SWAP} gate using
only generalised {\small CNOTs}. Finally, our understanding of the
symmetric group and its decompositions may well be enhanced by
considering how to design quantum circuits to realise certain
subgroups of this group.

\ack

The author wishes to thank Prof. Peter Wild for many helpful
suggestions.

\section*{References}
\bibliography{Bib}

\end{document}